\documentstyle[pra,multicol,aps,amstex,epsfig]{revtex}
\begin{document}

\setlength{\textwidth}{150mm}
\setlength{\textheight}{240mm}
\setlength{\parskip}{2mm}

\input{epsf.tex}
\epsfverbosetrue

\draft

\renewcommand{\baselinestretch}{1.0}

\title{Collapse arrest and soliton stabilization in nonlocal nonlinear media}

\author{Ole Bang$^1$, Wieslaw Krolikowski$^2$, John Wyller$^3$, 
        Jens Juul Rasmussen$^4$}

\address{$^1$ Department of Informatics and Mathematical Modelling, Technical 
         University of Denmark, DK-2800 Kgs.~Lyngby, Denmark.\\
         $^2$ Australian Photonics Cooperative Research Centre,
         Laser Physics Centre, Research School of Physical \\
         Sciences and Engineering, Australian National University, 
         Canberra ACT 0200, Australia.\\
         $^3$ Department of Mathematical Sciences,
         Agricultural University of Norway, P.O. Box 5035,
         N-1432 {\AA}s, Norway. \\
         $^4$Ris{\o} National Laboratory, Optics and Fluid Dynamics 
         Department, OFD-128 P.O.~Box 49, DK-4000 Roskilde, Denmark.}

\maketitle


\begin{abstract}
We investigate the properties of localized waves in systems governed by 
nonlocal nonlinear Schr\"{o}dinger type equations. 
We prove {\em rigorously} by bounding the Hamiltonian that nonlocality 
of the nonlinearity prevents collapse in, e.g., Bose-Einstein condensates 
and optical Kerr media in all physical dimensions.
The nonlocal nonlinear response must be symmetric, but can be of 
{\em completely arbitrary shape}.
We use variational techniques to find the soliton solutions 
and illustrate the stabilizing effect of nonlocality.
\end{abstract}


\begin{multicols}{2}

\narrowtext

Collapse is a fundamental physical phenomenon well-known in the theory of 
waves in nonlinear media. 
It refers to the situation when strong contraction or self-focusing of a 
wave leads to a catastrophic increase or blow-up of its amplitude after 
a finite time or propagation distance (see 
\cite{RasRyp86,Berge98,KivPel00} for reviews). 
Wave-collapse has been observed in plasma waves \cite{plasma-collapse}
(the famous Langmuir wave-collapse), electromagnetic waves or laser beams
\cite{laser-collapse} (also called self-focusing), Bose-Einstein Condensates 
(BEC's) or matter waves \cite{BEC-collapse}, and even capillary-gravity waves 
on deep water \cite{deepwater-collapse}.
The effect of collapse appears also in astrophysics, where the gravitational 
attraction plays the same role as the self-focusing of electromagnetic waves, 
tending to compress stars of sufficient mass, eventually leading to their
collapse into a black hole \cite{astro-collapse}.

Typically the contraction must be able to act freely in two or more 
dimensions to be strong enough to generate a collapse.
Moreover, the so-called norm, which is the power for electromagnetic and 
plasma waves, the atom density for BEC's, and the mass for stars, must
be above a certain critical value for a collapse to occur.
Most commonly the collapse has been discussed in the context of the 
nonlinear Schr\"odinger (NLS) equation, which is a universal model for 
dispersive (or diffractive) weakly nonlinear physical systems \cite{SulSul99}.
The NLS equation models, e.g., all systems mentioned above, in which 
a wave-collapse has been predicted and verified experimentally.

The collapse singularity is an artifact of the model and signals the limit 
of its validity.
Close to the singularity, when the amplitude is extremely high and the 
temporal- and spatial scales are extremely short, new physical processes will 
come into play \cite{RasRyp86,Berge98,KivPel00}. 
A common effect is nonlinear dissipation, such as two-photon absorption
of electromagnetic waves and inelastic two- and three-body recombination 
for matter waves, which efficiently absorbs the collapsing part of the 
wave. 
Thus collapse acts as an effective nonlinear loss mechanism, as is 
well-known in, e.g., Langmuir turbulence \cite{plasmareview} and BEC's 
\cite{BECreview}.
Effects, such as discreteness, saturation of the nonlinearity, and 
nonparaxiality, will also eliminate the possibility of a collapse
(see \cite{KivPel00}).
In contrast, effects such as weak linear loss \cite{KivPel00}, 
temperature fluctuations \cite{Bang-noise}, and spatial incoherence 
\cite{BanEdmKro99}, cannot eliminate the collapse.
In any case the collapse effect represents a strong mechanism for energy
localization, which is important to study to understand the properties
of a given physical system.

The inherent nonlocal character of the nonlinearity has attracted 
considerable interest as a means of eliminating collapse and stabilizing 
multidimensional solitary waves. 
Nonlocality appears naturally in optical systems with a thermal 
\cite{thermal} or diffusive \cite{SutBla93} type of nonlinearity.
Nonlocality is also known to influence the propagation of electromagnetic 
waves in plasmas \cite{litvak,DavFis95,juul,Berge00} and plays 
an important role in the theory of BEC's, where it accounts for the
finite-range many-body interaction 
\cite{BECreview,Goral00,PerKonGar00,ParSalRea98}. 

In this work we consider NLS equations with a general nonlocal
form of the nonlinearity. Turitsyn proved the absence of collapse for 
three {\em particular shapes} of the nonlocal nonlinear response 
\cite{Tur85}. 
The analysis of the collapse conditions for general response 
functions is difficult and has been carried out only numerically 
\cite{PerKonGar00}.
However, in many systems, such as BEC's, one has no knowledge of
the particular response function, and thus it is important to maintain
its generality in the model.
Here we prove {\em rigorously} that nonlocality eliminates
collapse in {\em all physical dimensions for arbitrary shapes} of 
the nonlocal response, as long as it is symmetric.  

We consider the evolution of a wave field $u=u(\vec{r})=u(\vec{r},\tau)$  
described by the general nonlocal NLS equation 
\begin{equation}
  \label{dynam}
  i\frac{\partial u}{\partial\tau} + \nabla^2u - V(\vec{r})u + N(I)u = 0, 
\end{equation}
where $V$=$V(\vec{r}\,)$ is an external (linear) confining potential, 
$I$=$I(\vec{r}\,)$=$I(\vec{r},\tau)$=$|u|^2$, $\tau$ is the evolution 
coordinate and $\vec{r}$=$(r_1,r_2,r_3)$ spans a D-dimensional "transverse"
coordinate space. 
The nonlinear term $N$=$N(I)$ is represented in the general nonlocal form 
\begin{equation}
  N(I) = \int R(\vec{r}^{\,\prime}-\vec{r}\,)I(\vec{r}^{\,\prime})
  d\vec{r}^{\,\prime},
  \label{N}
\end{equation}
where we assume the response function $R(\vec{r}\,)$ to be real, localized, 
nonsingular, and symmetric. 
For localized waves Eq.~(\ref{dynam}) conserve the power (in 
optics) or number of atoms (for BEC) $P$ and the Hamiltonian $H$,
\begin{equation}
  P = \int I d\vec{r}, \;\;
  H = \int[|\nabla u|^2+VI-\frac{NI}{2}]d\vec{r}.
\end{equation}
In optics $u$ is the envelope of the electric field with intensity 
$I$ and $V$ represents a guiding structure (waveguide).
Here Eq.~(\ref{N}) represents a general phenomenological model for 
self-focusing Kerr-like media, in which the change in the refractive 
index induced by an optical beam involves a transport process. 
This includes heat conduction in materials with a thermal nonlinearity 
\cite{thermal} or diffusion of molecules or atoms in atomic vapours 
\cite{SutBla93}.
A nonlocal response in the form (\ref{N}) appears naturally due
to many body interaction processes in the description of BEC's 
\cite{BECreview,PerKonGar00,ParSalRea98,FedJen01}, if the assumption of a 
zero-range interaction-potential is relaxed \cite{FedJen01}.
For BEC's with a negative scattering length
Eq.~(\ref{dynam}) is the nonlocal Gross-Pitaevskii (G-P) equation for the
collective wave function $u$ ($\tau$ is time), with $I$ representing the 
density of atoms and $V$ representing the magnetic trap.

In the limit when the response function is a delta-function, 
$R(\vec{r})$=$\delta(|\vec{r}|)$, the nonlinear response
is local (see Fig.~1a) and simply given by
\begin{equation}
  N(I) = I,
  \label{local}
\end{equation}
as in local optical Kerr media described by the standard NLS equation 
and in BEC's described by the standard G-P equation.
In this local limit multidimensional optical beams with a power higher 
than a certain critical value ($P>11.69$) will experience unbounded 
self-focusing and {\em collapse} after a finite propagation distance 
\cite{RasRyp86,Berge98,KivPel00}.
It is also well-known that BEC's would collapse when the total number
of atoms is larger than a critical number \cite{BECreview}.

With increasing width of the response function $R(\vec{r})$ the
wave intensity in the vicinity of the point $\vec{r}$ also contributes
to the nonlinear response at that point.
In case of weak nonlocality, when $R(\vec{r})$ is much narrower than
the width of the beam (see Fig.~1b), one can expand $I(\vec{r}^{\,\prime})$
around $\vec{r}^{\,\prime}$=$\vec{r}$ and obtain the simplified model
\begin{equation}
  \label{Nweak}
  N(I) = I + \gamma\nabla^2I, \quad 
  \gamma = \frac{1}{2}\int r^2 R(r) d\vec{r},
\end{equation}
where the small positive definite parameter $\gamma$ measures the
relative width of the nonlocal response.
The diffusion type model (\ref{Nweak}) of the nonlocal nonlinearity is a
model in its own right in plasma physics, where $\gamma$ can take any sign
\cite{litvak,DavFis95}.
It was also applied to BEC's \cite{ParSalRea98}, nonlinear optics
\cite{KroBan01}, and energy transfer in biomolecules \cite{molecular}.
In weakly nonlocal media with $N(I) = I + \gamma\nabla^2I$ it is
straightforward to show that collapse cannot occur.
This was first done for plasmas \cite{DavFis95}, and later for BEC's 
\cite{ParSalRea98}.

\begin{figure}
  \hspace{5mm}\epsfig{file=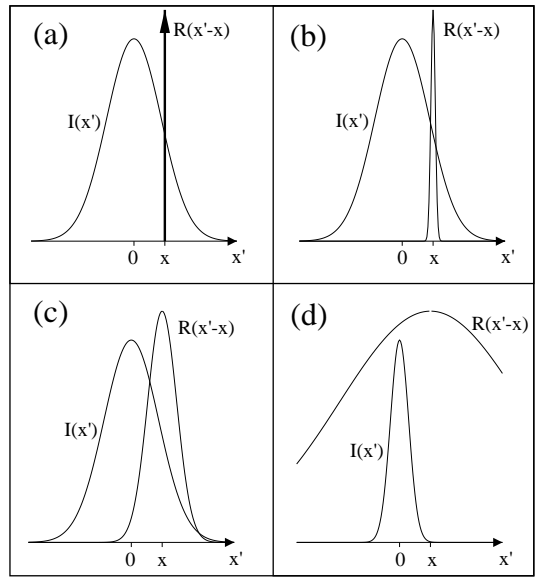,width=7cm, angle=0}
  \vspace{1mm}
  \caption{Degrees of nonlocality, as given by the relative width of the 
  response function $R$ and the intensity profile $I$ in the $x$-plane.
  Shown is the local (a), the weakly nonlocal (b), the general (c),
  and the strongly nonlocal (d) response.}
  \label{cases}
\end{figure}

In the limit of a strongly nonlocal response much broader than the
characteristic width of the wave function (see Fig.~1d), one can instead
expand the response function and obtain (to lowest order) 
\begin{equation}
   N(I) \approx P(R_0 + R_2r^2).
 \label{Nlinear}
\end{equation}
where $R_0=R(\vec{0})$ and $R_2=\frac{1}{2}\nabla^2R(\vec{0})$.
The evolution of optical beams in such a 
strongly nonlocal medium was considered in \cite{SnyMit97}.
Since this relation is linear, the highly nonlinear effect of 
collapse cannot occur.  

So in the two extreme limits of a weakly and highly nonlocal nonlinear
response the collapse is prevented.
For arbitrary degree of nonlocality it is difficult to prove anything
rigorously. Just saying that the dynamics is described by 
either the weakly nonlocal model (\ref{Nweak}) or the linear oscillator 
model (\ref{Nlinear}), which both have no collapse, is not enough.
As it is well-known from studies of general NLS equations, the typical
singularity is a so-called blow-up featuring the amplitude locally going
to infinity on a broad background localized structure \cite{RasRyp86}.
Such a two-scale field distribution, which was also recently observed in 
BEC's \cite{PerKonGar00}, is clearly described by neither of the two simple 
limiting systems.

The stabilizing effect of nonlocality of an arbitrary degree was proven 
by Turitsyn for three specific examples, including Coulomb interaction 
($R(\vec{r})$=$1/|\vec{r}|$) \cite{Tur85}.
He bounded the Hamiltonian from below for fixed power, which proves that a
collapse cannot occur and that the soliton solutions are stable in a 
Lyapunov sence.
Here we consider the general case of {\em arbitrarily shaped, nonsingular 
response functions} and prove rigorously that the Hamiltonian is bounded 
from below in all dimensions. 

Introducing the D-dimensional Fourier transform (denoted with a tilde)
and its inverse
\begin{equation}
  \tilde{I}(\vec{k}\,) = \int I(\vec{r}\,)
                       {\rm e}^{i\vec{k}\cdot\vec{r}}d\vec{r}, \quad
  I(\vec{r}\,) = \frac{1}{(2\pi)^D} \int \tilde{I}(\vec{k}\,)
               {\rm e}^{-i\vec{k}\cdot\vec{r}}d\vec{k},
\end{equation}
it is straightforward to show that for $N(I)$ given by Eq.~(\ref{N})
the following relations hold
\begin{eqnarray}
  & & |I(\vec{k})| = \left |\int I(\vec{r}){\rm e}^{i\vec{k}\cdot\vec{r}}
      d\vec{r} \,\right | \le \int I d\vec{r} = P, \\
  & & \int NId\vec{r}  =   \frac{1}{(2\pi)^D} \int \tilde{R}(\vec{k}\,)
      |\tilde{I}(\vec{k}\,)|^2d\vec{k}.  
\end{eqnarray} 
For response functions with a positive definite spectrum, $\tilde{R}
(\vec{k}\,)\ge0$, the Hamiltonian is thus bounded as follows
\begin{equation}
   H \ge ||\nabla u||_2^2 - \frac{1}{2} R_0 P^2,
   \label{Hbound}
\end{equation}
where $||u||_p^p\equiv\int|u|^pd\vec{r}$ and we have used the fact
that the minimum of $V(\vec{r}\,)$ is zero.

The inequality (\ref{Hbound}) is the main result of
this letter. It shows that for all symmetric response functions with a 
positive definite Fourier spectrum and a finite value at the center, the 
Hamiltonian is bounded from below by the conserved quantity -$\frac{1}{2} 
R_0 P^2$, or conversely, that the gradient norm $||\nabla u||_2^2$ is 
bounded from above by the conserved quantity $H+\frac{1}{2}R_0P^2$.
{\em This represents a rigorous proof that a collapse with the wave-amplitude
locally going to infinity cannot occur in BEC's, plasma, or optical Kerr media
with a nonlocal nonlinear response, for any physically reasonable
response function.}

The stabilising effect of the nonlocality can be further illustrated
by the properties of the stationary solutions of Eqs.~(\ref{dynam}). 
As a simple example we consider nonlocal optical bulk Kerr media 
with a Gaussian response
\begin{equation}
  \label{Rgaus}
  R(\vec{r}^{\,\prime}-\vec{r}\,) = \left( \frac{1}{\pi\sigma^2} \right)
  ^{\hspace{-1mm}\frac{D}{2}} 
  \exp\left(-\frac{|\vec{r}^{\,\prime}-\vec{r}\,|^2}{\sigma^2}\right).
\end{equation}
The ground-state stationary solutions are then radially symmetric
bell-shaped, nodeless solutions of the form $u(\vec{r},z) = \phi(r)
\exp(i\lambda z)$, where the profile $\phi(r)$ is found from the 
Euler-Lagrange equations for the Lagrangian
\begin{equation}
  L = \int \left[ \lambda\phi^2 + |\nabla \phi|^2 -
      \frac{1}{2}N(\phi^2)\phi^2 \right] d\vec{r}.
  \label{L}
\end{equation}
To capture the main physical effects we use the approximate
variational technique with a Gaussian trial profile $\phi(r)=\alpha\exp
[-(r/\beta)^2]$, in view of the fact that the Gaussian profile is an exact
solution in the strongly nonlocal limit with $N(I)$ given by the 
parabolic potential (\ref{Nlinear}).

\begin{figure}
  \setlength{\epsfxsize}{9cm}
  \centerline{\mbox{\epsffile{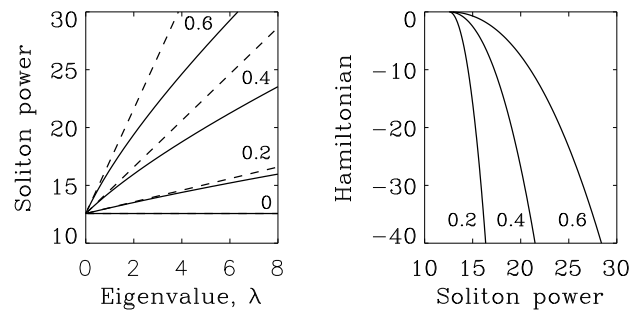}}}
  \vspace{2mm}
  \caption{2D variational results with Gaussian response and trial function.
  Left: Soliton power (solid) versus eigenvalue $\lambda$ for different 
  degrees of nonlocality, $\sigma$=0, 0.2, 0.4, and 0.6. Dashed lines 
  show the weakly nonlocal approximation.
  Right: Corresponding Hamiltonian versus power diagrams.}
  \label{2Dsol}
\end{figure}

Inserting this ansatz into the Lagrangian (\ref{L}), with $N$ given by the
general expression (\ref{N}), the Euler-Lagrange equations give the 
amplitude $\alpha$ and width $\beta$
\begin{eqnarray}
  \alpha^2 &=& (\lambda+D/\beta^2)(2+2\sigma^2/\beta^2)^{D/2}, \nonumber \\
  \beta^2  &=& [4-D+\sqrt{(4-D)^2+16\lambda\sigma^2}]/(2\lambda), 
\end{eqnarray}
In Fig.~\ref{2Dsol} we show the power $P_s$=$(\pi/2)^{D/2}\alpha^2\beta^D$
and Hamiltonian of the stationary 
solutions in 2D. The dashed lines give the results of the weakly nonlocal 
approximation with $N$ given by Eq.~(\ref{Nweak}), from which 
$\alpha^2=4\lambda$ and $\beta^2=2/\lambda+2\sigma^2$ is found, 
resulting in the power
\begin{equation}
  \label{Psweak}
  P_s   = 4\pi(1+\sigma^2\lambda),
\end{equation}
where $4\pi$ is the ($\lambda$-independent) power of the Gaussian 
approximation to the soliton solution of the standard 2D NLS equation,
recovered in the local limit $\sigma$=0.

In the 2D NLS equation the collapse is critical and the stationary 
solutions are "marginally stable" with $dP_s/d\lambda$=0 
\cite{RasRyp86,Berge98,SulSul99}.
Typically perturbations act against the self-focusing, with several 
effects, such as non-paraxiality and saturability, completely eliminating
collapse \cite{KivPel00}.
This is also the case with nonlocality, as evidenced from Fig.~\ref{2Dsol}
and the simplified expression (\ref{Psweak}), which shows that any
finite width of the response function (non-zero value of $\sigma$) 
implies that $dP_s/d\lambda$ becomes positive definite.
According to the (necessary) Vakhitov-Kolokolov (VK)   
criterion \cite{Vakhitov} the soliton 
solutions therefore (possibly) become {\em linearly stable}. 

For small $\lambda$ the soliton width $\beta$ decreases as $1/\lambda$. 
Thus the assumption of weak nonlocality, i.e. that the soliton is much 
wider than the response function, applies only to sufficiently small 
values of $\lambda$ satisfying $\lambda\sigma^2\ll1$, which is also 
clearly seen from Fig.~\ref{2Dsol}. 
The accuracy of the assumption of weak nonlocality is further discussed 
in Ref.~\cite{KroBanJuuWyl01} in terms of modulational instability. 

\begin{figure}
  \setlength{\epsfxsize}{9cm}
  \centerline{\mbox{\epsffile{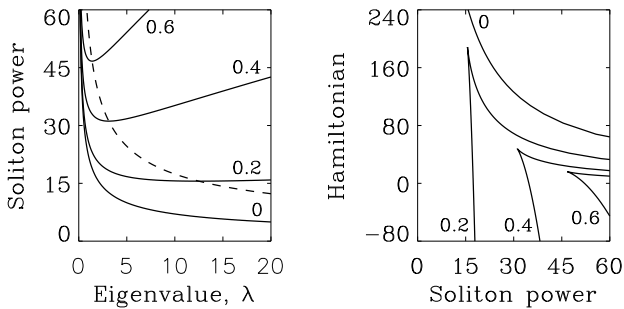}}}
  \vspace{2mm}
  \caption{3D variational results with Gaussian response and trial function.
  Left: Soliton power (solid) versus eigenvalue $\lambda$ for different 
  degrees of nonlocality, $\sigma$=0, 0.2, 0.4, and 0.6. 
  Dashed lines show the threshold power (\ref{threshold}).
  Right: Corresponding Hamiltonian versus power diagrams.}
\label{3Dsol}
\end{figure}

The 3D case shown in Fig.~\ref{3Dsol} is more interesting, because the
nonlinearity is much stronger than in 2D. The collapse in the local 3D NLS
equation is so-called supercritical \cite{RasRyp86,Berge98,SulSul99}.
Again the soliton width $\beta$ decreases as $1/\lambda$, so a threshold 
width should exist, below which the nonlocality is not strong 
enough to stabilize the soliton. 
This is exactly what is observed in Fig.~\ref{3Dsol}: For $\lambda<
\lambda^{\rm th}$ the solitons are still linearly unstable with 
$dP_s/d\lambda<0$, but above threshold the nonlocality is strong 
enough to bend the curve and make $dP_s/d\lambda>0$, i.e., the solitons
become linearly stable according to the VK criterion. 
From the definition $dP_s(\lambda^{\rm th})/d\lambda$=0 the variational 
results give $\lambda^{\rm th}=1/(2\sigma^2)$, corresponding to a 
threshold in the soliton power (dashed curve in Fig.~\ref{3Dsol}) and width
\begin{equation}
   P_s^{\rm th} = (5\pi)^{3/2}5\sigma/4, \quad
   \beta^{\rm th} = 2\sigma,
   \label{threshold}
\end{equation}
which are both proportional to the degree of nonlocality $\sigma$.
Thus, sufficiently broad and high-power solitons are stable. 
In the Hamiltonian versus power diagram in Fig.~\ref{3Dsol} the lower 
(upper) branches correspond to stable (unstable) solutions while the 
threshold is represented by the cusp \cite{AkhAnkGri99}. 
This stable solution branch was recently found numerically in the context 
of BEC's with a nonlocal negative scattering potential \cite{ParSalRea98}. 
It corresponds to high density, self bound states of the condensate. 

In conclusion we studied the properties of localized wave packets in 
nonlocal NLS equations.  We proved {\em rigorously} that nonlocality 
of an {\em arbitrary shape} eliminates collapse in {\em all physical 
dimensions}. We also demonstrated that multidimensional soliton 
solutions of the NLS equation may be stabilized by the nonlocality.
This opens a new and interesting discussion as to what is actually 
observed in collapse experiments in nonlocal systems.
It seems clear that it all comes down to oscillations between opposite
extreme states and how strong and rapid they are.
Such oscillations were recently found to occur in BEC's through numerical
and variational studies \cite{PerKonGar00}.

We acknowledge useful discussions with Yu.S. Kivshar, S.K. Turitsyn, and
Yu.B. Gaididei.  This work was supported by the Danish Technical Research 
Council (Talent Grant No.~26-00-0355) and the Danish Natural Sciences
Foundation (Grant No.~9903273).

\end{multicols}

\end{document}